\documentclass[12pt]{iopart}
\usepackage{iopams} 
\usepackage{mathbbol} 
\usepackage[colorlinks, linkcolor =blue]{hyperref} 
\usepackage{amsthm} 
\usepackage{mathrsfs} 
\usepackage{array} 
\usepackage{multirow} 
\usepackage{color}
\usepackage{ulem}

\newcommand{\bm}[1]{\boldsymbol{#1}} 
\newtheorem*{lemma*}{Lemma} 
\newtheorem*{proposition*}{Proposition} 

\newcolumntype{M}[1]{>{\centering\arraybackslash}m{#1}}
\newcolumntype{N}{@{}m{0pt}@{}}
\newcolumntype{?}{!{\vrule width 3pt}} 

\begin{document}


\title{Microreversibility, nonequilibrium response, and Euler's polynomials}

\author{M Barbier and P Gaspard}

\address{Center for Nonlinear Phenomena and Complex Systems, Universit\'e Libre de Bruxelles (ULB), Code Postal 231, Campus Plaine, B-1050 Brussels, Belgium}

\ead{Maximilien.Barbier@ulb.ac.be,gaspard@ulb.ac.be}


\begin{abstract}
Microreversibility constrains the fluctuations of the nonequilibrium currents that cross an open system. This can be seen from the so-called fluctuation relations, which are a direct consequence of microreversibility. Indeed, the latter are known to impose time-reversal symmetry relations on the statistical cumulants of the currents and their responses at arbitrary orders in the deviations from equilibrium. Remarkably, such relations have been recently analyzed by means of Euler's polynomials. Here we show that fluctuation relations can actually be explicitly written in terms of the constant terms of these particular polynomials. We hence demonstrate that Euler's polynomials are indeed fundamentally rooted in fluctuation relations, both in the absence and the presence of an external magnetic field.
\end{abstract}

\vspace{2pc}
\noindent{\it Keywords}: Euler's polynomials, time-reversal symmetry, fluctuation relations, nonequilibrium systems, response theory.


\section{Introduction}\label{intro_sec}

Microreversibility expresses the symmetry of the microscopic Hamiltonian dynamics of a system under the time-reversal transformation \cite{deGr}. Such a symmetry holds both for classical and quantum systems in the absence or the presence of an external magnetic field $\bm{B}$. In the latter case, the time-reversal symmetry applies to the total system that includes the charged particles in the external circuits whose electric currents generate the magnetic field $\bm{B}$. This symmetry lies at the heart of the study of nonequilibrium systems, as a large number of important results of nonequilibrium statistical physics appear to be fundamentally rooted in microreversibility.

Nonequilibrium systems are characterized by the occurrence of net currents, e.g., of energy or particles. The latter describe the response of the system to the nonequilibrium constraints to which it is subjected. Such constraints can be mechanical or thermodynamic forces, commonly referred to as affinities \cite{deGr,deDon,Prig} and rising, in particular, from differences of temperatures or chemical potentials. Close to equilibrium, the currents are proportional to the affinities. This is for instance the case in Fick's law of diffusion \cite{deGr} or Ohm's law \cite{Imry}. In this linear response regime, microreversibility manifests itself into the well-known Onsager-Casimir reciprocity relations satisfied by the linear response coefficients \cite{Ons31a,Ons31b,Cas45}, the Green-Kubo formulae \cite{Gre52,Gre54,Kub57}, or the fluctuation-dissipation theorem \cite{CW51,Kub66}.

However, many nonequilibrium systems operate in regimes where the currents have nonlinear dependences on the nonequilibrium constraints. This is for instance the case in mesoscopic electronic circuits where large voltage differences can be implemented, hence inducing deviations from Ohm's law \cite{NYH10,NYH11}. Remarkably, microreversibility greatly influences the nonlinear transport properties of a nonequilibrium system as well, as was first noted for Hamiltonian systems or stochastic processes \cite{BK77,S92,S94}. The more recent \textit{fluctuation relations} (or fluctuation theorems) are also deeply rooted in microreversibility \cite{ECM93,ES94,GC95,Jar97,Kur98,LS99,Cro99,ZC03,Kur00,AGM09,EHM09,CHT11,HPPG11,Gas13_1, Gas13_2}. The latter stand as exact results that remain valid arbitrarily far from equilibrium, and typically express a particular symmetry relation satisfied by the probability distribution of the nonequilibrium currents.

We recently studied the fluctuation relation for open systems, connected to reservoirs of energy and particles, in nonequilibrium steady states both in the absence \cite{BG18} and the presence \cite{BG19} of an external magnetic field $\bm{B}$. We showed that the fluctuation relation reduces by about half the total number of statistical cumulants, and of their responses to the affinities, that need to be known in order to fully describe the statistical properties of the nonequilibrium currents. Interestingly, the analysis performed in \cite{BG18,BG19} made extensive use of Euler's polynomials (see e.g. \cite{AbrSteg,GradRyz}). In particular, we showed in \cite{BG19} that the cumulants that are constrained by the fluctuation relation can be expressed as linear combinations of the unconstrained ones, the coefficients of which being precisely the coefficients of Euler's polynomials. Surprisingly however, the latter have not been previously of much use in nonequilibrium statistical physics. To further investigate the connection between Euler's polynomials and fluctuation relations, one of the most important quantitative tool of nonequilibrium statistical physics, is thus an important question of theoretical interest.

In this paper, we demonstrate that Euler's polynomials are fundamentally rooted in the fluctuation relation, and this for systems both in the absence and the presence of external magnetic fields. In our case, the fluctuation relation consists in a symmetry property satisfied by the generating function of the statistical cumulants of the nonequilibrium currents. We show that it can be adequately rewritten in a form that explicitly involves the coefficients of Euler's polynomials. This alternative expression of the fluctuation relation allows us to readily express some of the cumulants and their responses as linear combinations of the remaining ones, and thus to unambiguously recover results that we previously obtained in \cite{BG19}. By introducing Euler's polynomials at the level of the fluctuation relation itself, the present work hence clarifies the fundamental connection, touched upon in \cite{BG18,BG19}, between these particular polynomials and microreversibility.

We first state in section~\ref{FR_sec} the fluctuation relation that we consider throughout this work, before we discuss in section~\ref{Euler_sec} some of the properties of Euler's polynomials. We then reformulate in section~\ref{alt_FR} the fluctuation relation by means of Euler's polynomials, and introduce the statistical cumulants in section~\ref{cumul_sec}. Here we show that the alternative form of the fluctuation relation obtained in section~\ref{alt_FR} readily yields results previously obtained in \cite{BG19}. Concluding remarks are drawn in section~\ref{conclusion_sec}.


\section{Fluctuation relation}\label{FR_sec}

We consider an open system connected to reservoirs of energy and particles in the presence of an external magnetic field $\bm{B}$. The system is assumed to reach a nonequilibrium steady state after a long enough time. The statistical properties of the nonequilibrium currents that cross the system can then be described by the generating function $Q \left( \bm{\lambda} , \bm{A} ; \bm{B} \right)$ of the statistical cumulants. It is a function of the counting parameters $\bm{\lambda}$ associated with the currents, the affinities $\bm{A}$ that drive them away from equilibrium, and the magnetic field (the latter being treated as a parameter in the sequel). It is worth specifying that the dimension of the vectors $\bm{\lambda}$ and $\bm{A}$ is equal to the total number of currents. The function $Q$ is known \cite{AGM09,Gas13_1,Gas13_2,SU08} to satisfy the multivariate fluctuation relation
\begin{eqnarray}
Q \left( \bm{\lambda} , \bm{A} ; \bm{B} \right) = Q \left( \bm{A} - \bm{\lambda} , \bm{A} ; -\bm{B} \right)
\label{fluct_rel}
\end{eqnarray}
as a consequence of microreversibility.

After the substitution $\bm{\lambda}\to -\bm{\lambda}$, and assuming the generating function $Q$ to be analytic so that the action of the derivatives $\partial_{\bm{\lambda}} \equiv \partial / \partial \bm{\lambda}$ is well defined, the fluctuation relation~\eref{fluct_rel} can be written as
\begin{eqnarray}
Q \left( -\bm{\lambda} , \bm{A} ; \bm{B} \right) = {\rm e}^{\bm{A}\cdot\partial_{\bm{\lambda}}} \, Q \left( \bm{\lambda} , \bm{A} ; -\bm{B} \right)
\label{fluct_rel_alt}
\end{eqnarray}
in terms of the translation operator ${\rm e}^{\bm{A}\cdot\partial_{\bm{\lambda}}}$, acting as ${\rm e}^{\bm{A}\cdot\partial_{\bm{\lambda}}}g(\bm{\lambda})=g(\bm{\lambda}+\bm{A})$ on any function $g(\bm{\lambda})$.


\section{Euler's polynomials}\label{Euler_sec}

The generating function of Euler's polynomials $E_n(x)$ has the form
\begin{eqnarray}
\frac{2\, {\rm e}^{xt}}{{\rm e}^t+1} = \sum_{n=0}^{\infty} \frac{1}{n!} \, E_n(x) \, t^n
\label{Euler_poly_GF}
\end{eqnarray}
for $x$ a real number \cite{AbrSteg,GradRyz}, where the $n^{\mathrm{th}}$ order polynomial $E_n(x)$ can be written in the form
\begin{eqnarray}
E_n(x) = \sum_{i=0}^{n} e_i^{(n)} \, x^i \, .
\label{Euler_poly_def}
\end{eqnarray}
Taking $x=0$ in \eref{Euler_poly_GF} gives
\begin{eqnarray}
\frac{2}{{\rm e}^t+1} = \sum_{n=0}^{\infty} \frac{1}{n!}  \, e_0^{(n)}  \,  t^n \, ,
\label{Euler_poly_GF_0}
\end{eqnarray}
hence generating the coefficients
\begin{eqnarray}
e_0^{(n)} \equiv E_n(0) \, .
\label{e0n}
\end{eqnarray}
Setting $x=1$ in \eref{Euler_poly_GF}, we instead get that
\begin{eqnarray}
\frac{2\, {\rm e}^t}{{\rm e}^t+1} = \sum_{n=0}^{\infty} \frac{1}{n!}  \, E_n(1)  \,  t^n \, .
\label{Euler_poly_GF_1}
\end{eqnarray}
Since
\begin{eqnarray}
\frac{2\, {\rm e}^{t}}{{\rm e}^t+1} = \frac{2}{{\rm e}^{-t}+1} = \sum_{n=0}^{\infty}  \frac{(-1)^n}{n!}  \, e_0^{(n)} \, t^n  \, ,
\end{eqnarray}
we find that
\begin{eqnarray}
e_0^{(n)}= E_n (0) = (-1)^n \, E_n(1) \, .
\label{e0n_0+1}
\end{eqnarray}
Adding together~\eref{Euler_poly_GF_0} and~\eref{Euler_poly_GF_1} leads to
\begin{eqnarray}
\sum_{n=0}^{\infty}  \frac{1}{n!}  \,  [E_n(0)+E_n(1)]  \, t^n = 2 \, ,
\label{Euler_nbers_GF_0+1}
\end{eqnarray}
so that 
\begin{eqnarray}
E_0(0)+E_0(1) =2 \, , \qquad\mbox{and} \qquad 
E_n (0) + E_n(1) = 0 \quad\mbox{for} \quad n\ge 1 \, .
\label{Euler_nbers_rel}
\end{eqnarray}
As a consequence of~\eref{e0n_0+1} and~\eref{Euler_nbers_rel}, we recover the properties that
\begin{eqnarray}
&& e_0^{(0)}=E_0(0)= E_0(1)= 1 \, , 
\end{eqnarray}
and
\begin{eqnarray}
&& e_{0}^{(2j-1)} = E_{2j-1}(0)= - E_{2j-1}(1)   \, , \\
&& e_{0}^{(2j)} = E_{2j}(0)= E_{2j}(1) = 0 \, , 
\end{eqnarray}
for $j\ge 1$ \cite{AbrSteg,GradRyz}.

Now, since the hyperbolic tangent can be expressed as
\begin{eqnarray}
\tanh \frac{t}{2} = \frac{{\rm e}^{t/2}-{\rm e}^{-t/2}}{{\rm e}^{t/2}+{\rm e}^{-t/2}} = \frac{{\rm e}^{t}-1}{{\rm e}^{t}+1} = \frac{1}{{\rm e}^{-t}+1} - \frac{1}{{\rm e}^t+1}  \, ,
\label{tanh}
\end{eqnarray}
we find the following relation between the hyperbolic tangent and the constant terms of Euler's polynomials:
\begin{eqnarray}
\tanh \frac{t}{2} = -   \sum_{j=1}^{\infty}  \frac{e_0^{(2j-1)}}{(2j-1)!} \, t^{2j-1}
  \, .
\label{tanh-e0n}
\end{eqnarray}

It is also worth noting that the constant terms $e_0^{(n)}$ of Euler's polynomials are related to Bernoulli's numbers $B_n$ according to
\begin{eqnarray}
e_0^{(n)} = E_n(0) = - \frac{2}{n+1} \, (2^{n+1}-1) \, B_{n+1}  \, ,
\label{e0n-Bernoulli}
\end{eqnarray}
for any $n>0$, so that we recover the known power series expansion 
\begin{eqnarray}
\tanh \frac{t}{2} = \sum_{j=1}^{\infty}  c_{2j-1}\, t^{2j-1} \qquad \mbox{with}\qquad c_{2j-1} = \frac{2}{(2j)!} \, (2^{2j}-1) \, B_{2j}
\label{tanh-Bn}
\end{eqnarray}
(see e.g. equation 4.5.64 of reference \cite{AbrSteg}).


\section{Alternative forms of the fluctuation relation}\label{alt_FR}

We introduce the functions
\begin{eqnarray}
Q_{\pm}(\bm{\lambda},\bm{A};\bm{B}) \equiv \frac{1}{2} \left[ Q(\bm{\lambda},\bm{A};\bm{B})  \pm Q(-\bm{\lambda},\bm{A};\bm{B})  \right] \, ,
\label{Q_pm}
\end{eqnarray}
giving the parts of the cumulant generating function that are even ($Q_{+}$) or odd ($Q_{-}$) in the counting parameters $\bm{\lambda}$ and such that
\begin{eqnarray}
Q(\pm \bm{\lambda},\bm{A};\bm{B}) = Q_{+}(\bm{\lambda},\bm{A};\bm{B}) \pm Q_{-}(\bm{\lambda},\bm{A};\bm{B}) \, .
\end{eqnarray}
Besides, the symmetric and antisymmetric parts of an arbitrary function $f$ of $\bm{B}$ are defined as
\begin{eqnarray}
f^{\mathrm{S},\mathrm{A}}(\bm{B}) \equiv \frac{1}{2} \left[ f( \bm{B}) \pm f( -\bm{B}) \right] \, ,
\label{sym_antisym_gen_def}
\end{eqnarray}
so that
\begin{eqnarray}
f( \pm \bm{B}) = f^{\mathrm{S}}(\bm{B}) \pm f^{\mathrm{A}}(\bm{B}) \, .
\label{fct_from_sym_antisym}
\end{eqnarray}

Now, substituting the fluctuation relation~\eref{fluct_rel_alt} into the definition~\eref{Q_pm} of $Q_{\pm}$ yields
\begin{eqnarray}
Q_{\pm}(\bm{\lambda},\bm{A};\bm{B}) = \frac{1}{2} \left[ Q(\bm{\lambda},\bm{A};\bm{B})  \pm {\rm e}^{\bm{A}\cdot\partial_{\bm{\lambda}}} \, Q(\bm{\lambda},\bm{A};-\bm{B})  \right] \, .
\label{FR_Q_pm}
\end{eqnarray}
Moreover, taking the symmetric and antisymmetric parts of equations~\eref{FR_Q_pm} in the magnetic field $\bm{B}$ gives the four following relations:
\begin{eqnarray}
&& Q_{+}^{\mathrm{S}}(\bm{\lambda},\bm{A};\bm{B}) = \frac{1}{2} \left( 1 + {\rm e}^{\bm{A}\cdot\partial_{\bm{\lambda}}} \right) \, Q^{\mathrm{S}}(\bm{\lambda},\bm{A};\bm{B}) \, , \label{FR_Q+S} \\
&& Q_{-}^{\mathrm{S}}(\bm{\lambda},\bm{A};\bm{B}) = \frac{1}{2} \left( 1 - {\rm e}^{\bm{A}\cdot\partial_{\bm{\lambda}}} \right) \, Q^{\mathrm{S}}(\bm{\lambda},\bm{A};\bm{B}) \, , \label{FR_Q-S} \\
&& Q_{+}^{\mathrm{A}}(\bm{\lambda},\bm{A};\bm{B}) = \frac{1}{2} \left( 1 - {\rm e}^{\bm{A}\cdot\partial_{\bm{\lambda}}} \right) \, Q^{\mathrm{A}}(\bm{\lambda},\bm{A};\bm{B}) \, , \label{FR_Q+A}\\
&& Q_{-}^{\mathrm{A}}(\bm{\lambda},\bm{A};\bm{B}) = \frac{1}{2} \left( 1 + {\rm e}^{\bm{A}\cdot\partial_{\bm{\lambda}}} \right) \, Q^{\mathrm{A}}(\bm{\lambda},\bm{A};\bm{B}) \, . \label{FR_Q-A}
\end{eqnarray}
Multiplying~\eref{FR_Q+S} and~\eref{FR_Q-A} by $(1 - {\rm e}^{\bm{A}\cdot\partial_{\bm{\lambda}}})$, and \eref{FR_Q-S} as well as~\eref{FR_Q+A} by $(1 + {\rm e}^{\bm{A}\cdot\partial_{\bm{\lambda}}})$, shows that we have the identities
\begin{eqnarray}
&& \left( 1 - {\rm e}^{\bm{A}\cdot\partial_{\bm{\lambda}}} \right) \, Q_{+}^{\mathrm{S}}(\bm{\lambda},\bm{A};\bm{B}) = \left( 1 + {\rm e}^{\bm{A}\cdot\partial_{\bm{\lambda}}} \right) \, Q_{-}^{\mathrm{S}}(\bm{\lambda},\bm{A};\bm{B})\, , \label{FR_Q+-S} \\
&& \left( 1 + {\rm e}^{\bm{A}\cdot\partial_{\bm{\lambda}}} \right) \,Q_{+}^{\mathrm{A}}(\bm{\lambda},\bm{A};\bm{B}) = \left( 1 - {\rm e}^{\bm{A}\cdot\partial_{\bm{\lambda}}} \right) \, Q_{-}^{\mathrm{A}}(\bm{\lambda},\bm{A};\bm{B}) \, . \label{FR_Q+-A}
\end{eqnarray}
Inverting $(1 + {\rm e}^{\bm{A}\cdot\partial_{\bm{\lambda}}})$ and using~\eref{tanh}, these two relations equivalently read
\begin{eqnarray}
&& Q_{-}^{\mathrm{S}}(\bm{\lambda},\bm{A};\bm{B}) = - \tanh\left( \frac{1}{2}  \bm{A}\cdot\partial_{\bm{\lambda}} \right) \, Q_{+}^{\mathrm{S}}(\bm{\lambda},\bm{A};\bm{B})\, , \label{FR_Q+-S-th} \\
&& Q_{+}^{\mathrm{A}}(\bm{\lambda},\bm{A};\bm{B}) =  - \tanh\left( \frac{1}{2}  \bm{A}\cdot\partial_{\bm{\lambda}} \right) \, Q_{-}^{\mathrm{A}}(\bm{\lambda},\bm{A};\bm{B}) \, . \label{FR_Q+-A-th}
\end{eqnarray}
Finally, combining~\eref{FR_Q+-S-th}-\eref{FR_Q+-A-th} with the expansion~\eref{tanh-e0n} of the hyperbolic tangent in terms of the constant terms~\eref{e0n} of Euler's polynomials, we thus obtain
\begin{eqnarray}
&& Q_{-}^{\mathrm{S}}(\bm{\lambda},\bm{A};\bm{B}) = \sum_{j=1}^{\infty}  \frac{e_0^{(2j-1)}}{(2j-1)!}  \left( \bm{A}\cdot\partial_{\bm{\lambda}} \right)^{2j-1} \, Q_{+}^{\mathrm{S}}(\bm{\lambda},\bm{A};\bm{B})\, , \label{FR_Q+-S-Euler} \\
&& Q_{+}^{\mathrm{A}}(\bm{\lambda},\bm{A};\bm{B}) =  \sum_{j=1}^{\infty}  \frac{e_0^{(2j-1)}}{(2j-1)!}  \left( \bm{A}\cdot\partial_{\bm{\lambda}} \right)^{2j-1} \, Q_{-}^{\mathrm{A}}(\bm{\lambda},\bm{A};\bm{B}) \, , \label{FR_Q+-A-Euler}
\end{eqnarray}
which are equivalent to the original fluctuation relation~\eref{fluct_rel}.


\section{Cumulants and their responses}\label{cumul_sec}

The cumulants and their nonequilibrium responses around equilibrium are defined by
\begin{eqnarray}
Q_{\alpha_1 \cdots \alpha_m \, , \, \beta_1 \cdots \beta_n} (\bm{B}) \equiv \frac{\partial^{m+n} Q}{\partial \lambda_{\alpha_1} \cdots \partial \lambda_{\alpha_m} \partial A_{\beta_1} \cdots \partial A_{\beta_n}} \left( \bm{0}, \bm{0} ; \bm{B} \right) \, ,
\label{m_cumulant_n_resp_def}
\end{eqnarray}
and they can be obtained by expanding the cumulant generating function $Q$ in powers of the counting parameters $\bm{\lambda}$ and the affinities $\bm{A}$ according to
\begin{eqnarray}
\fl Q \left( \pm \bm{\lambda} , \bm{A} ; \bm{B} \right) = \sum_{m , n = 0}^{\infty} \frac{(\pm 1)^m}{m! \, n!} \, Q_{\alpha_1 \cdots \alpha_m \, , \, \beta_1 \cdots \beta_n} (\bm{B})\, \lambda_{\alpha_1} \cdots \lambda_{\alpha_m} A_{\beta_1} \cdots A_{\beta_n}
\label{Q_exp_count_par_and_aff}
\end{eqnarray}
with Einstein's convention of summation over repeated indices. The expansion~\eref{Q_exp_count_par_and_aff} readily allows us to obtain the corresponding power series of the functions $Q_{\pm}^{\mathrm{S},\mathrm{A}}$ by means of their definitions~\eref{Q_pm} and~\eref{sym_antisym_gen_def}.

Now, we have that
\begin{eqnarray}
\fl \left( \bm{A} \cdot \partial_{\bm{\lambda}} \right)^k Q \left( \pm \bm{\lambda} , \bm{A} ; \bm{B}\right) = \sum_{m=0}^{\infty}\sum_{n=k}^{\infty} \frac{(\pm 1)^{m+k}}{m! \, n!} \, Q_{\alpha_1 \cdots \alpha_m \, , \, \beta_1 \cdots \beta_n}^{\{k\}}(\bm{B}) \, \lambda_{\alpha_1} \cdots \lambda_{\alpha_m} A_{\beta_1} \cdots A_{\beta_n} \nonumber \\
\label{A_d_Q}
\end{eqnarray}
for any $k \geqslant 0$. In~\eref{A_d_Q}, the quantities $Q^{\{k\}}$ are defined by
\begin{eqnarray}
Q_{\alpha_1 \cdots \alpha_m \, , \, \beta_1 \cdots \beta_n}^{\{k\}}(\bm{B}) &\equiv& \sum_{j = 1}^{n} Q_{\alpha_1 \cdots \alpha_m \beta_j \, , \, \beta_1 \cdots \beta_{j-1} \beta_{j+1} \cdots \beta_n}^{\{k-1\}}(\bm{B}) \nonumber\\[0.2cm]
&=& \sum_{j_1 = 1}^{n} \sum_{j_{2}=1 \atop j_{2} \neq j_{1}}^{n} \cdots \sum_{j_{k}=1 \atop j_{k} \neq j_1 , \ldots , j_{k-1}}^{n} Q_{\alpha_1 \cdots \alpha_m \beta_{j_1} \cdots \beta_{j_k} \, , \, (\boldsymbol{\cdot})}(\bm{B})
\label{Q_k_def}
\end{eqnarray}
for $k \geqslant 1$, with $Q^{\{0\}} \equiv Q$, and where $(\boldsymbol{\cdot})$ denotes the set of all subscripts $\beta$ that are different from the subscripts $\beta$ present on the left of the comma, i.e., $\beta_{j_1} , \ldots , \beta_{j_k}$. The results~\eref{A_d_Q}-\eref{Q_k_def} can be shown by induction on the integer $k$ \cite{BG19}, and by noting that the differential operator $\bm{A} \cdot \partial_{\bm{\lambda}}  = A_{\gamma} \, \partial_{\lambda_{\gamma}}$ acts as 
\begin{eqnarray}
\fl \left( \bm{A} \cdot \partial_{\bm{\lambda}} \right) \, Q_{\alpha_1 \cdots \alpha_m \, , \, \beta_1 \cdots \beta_n}^{\{k\}} (\bm{B}) \, \lambda_{\alpha_1} \cdots \lambda_{\alpha_m} = m \, Q_{\alpha_1 \cdots \alpha_{m-1} \gamma \, , \, \beta_1 \cdots \beta_n}^{\{k\}} (\bm{B}) \, \lambda_{\alpha_1} \cdots \lambda_{\alpha_{m-1}} A_{\gamma} \, ,
\label{A_d_action_def}
\end{eqnarray}
where we used the invariance of the cumulants~\eref{m_cumulant_n_resp_def} under any permutation of the subscripts either on the left or the right of the comma and again Einstein's convention for repeated indices. One can thus see that the summation over the mute indices $\gamma,\beta_1, \ldots, \beta_n$ implied by Einstein's convention can be adequately rewritten (by means of a mere change of indices) so as to obtain
\begin{eqnarray}
\fl \left( \bm{A} \cdot \partial_{\bm{\lambda}} \right) \, Q_{\alpha_1 \cdots \alpha_m \, , \, \beta_1 \cdots \beta_n}^{\{k\}} (\bm{B}) \, \lambda_{\alpha_1} \cdots \lambda_{\alpha_m} A_{\beta_1} \cdots A_{\beta_n} \nonumber \\
= m \, Q_{\alpha_1 \cdots \alpha_{m-1} \beta_1 \, , \, \beta_2 \cdots \beta_{n+1}}^{\{k\}} (\bm{B}) \, \lambda_{\alpha_1} \cdots \lambda_{\alpha_{m-1}} A_{\beta_1} \cdots A_{\beta_{n+1}} \, .
\label{A_d_action_full}
\end{eqnarray}
The fact that the results~\eref{A_d_Q}-\eref{Q_k_def} remain true for the integer $k+1$ then readily follows from~\eref{A_d_action_full}.

In addition, it has been shown in reference~\cite{BG19} that
\begin{eqnarray}
Q_{\alpha_1 \cdots \alpha_m \, , \, \beta_1 \cdots \beta_n}^{\{k\}}(\bm{B}) = k! \, Q_{\alpha_1 \cdots \alpha_m \, , \, \beta_1 \cdots \beta_n}^{(k)}(\bm{B})
\label{Q_k_alt_expr}
\end{eqnarray}
in terms of the quantities
\begin{eqnarray}
Q_{\alpha_1 \cdots \alpha_m \, , \, \beta_1 \cdots \beta_n}^{(k)}(\bm{B}) \equiv \sum_{j_1 = 1}^{n} \sum_{j_{2}=1 \atop j_{2} > j_{1}}^{n} \cdots \sum_{j_{k}=1 \atop j_{k} > j_{k-1}}^{n} Q_{\alpha_1 \cdots \alpha_m \beta_{j_1} \cdots \beta_{j_k} \, , \, (\bm{\cdot})}(\bm{B})
\label{Q_k_expr}
\end{eqnarray}
for $k \geqslant 1$, with again $Q^{(0)} \equiv Q$.

Substituting the power series~\eref{Q_exp_count_par_and_aff}-\eref{A_d_Q} into~\eref{FR_Q+-S-Euler} and using~\eref{Q_k_alt_expr} generates the identity
\begin{eqnarray}
\fl \frac{1}{2} \sum_{m,n=0}^{\infty} \frac{1 - (- 1)^{m}}{m! \, n!} \, Q_{\alpha_1 \cdots \alpha_m \, , \, \beta_1 \cdots \beta_n}^{\mathrm{S}}(\bm{B}) \, \lambda_{\alpha_1} \cdots \lambda_{\alpha_m} A_{\beta_1} \cdots A_{\beta_n} \nonumber \\
\fl = \frac{1}{2} \sum_{m=0}^{\infty} \sum_{j=1}^{\infty} \sum_{n=2j-1}^{\infty} e_0^{(2j-1)} \, \frac{1 + (- 1)^{m+2j-1}}{m! \, n!} \, Q_{\alpha_1 \cdots \alpha_m \, , \, \beta_1 \cdots \beta_n}^{(2j-1) \, \mathrm{S}}(\bm{B}) \, \lambda_{\alpha_1} \cdots \lambda_{\alpha_m} A_{\beta_1} \cdots A_{\beta_n} \, .
\label{Q_min_S_Q_plus_S}
\end{eqnarray}
Identifying the terms with the same powers of $\lambda_{\alpha}$ and $A_{\beta}$ on both sides of~\eref{Q_min_S_Q_plus_S}, and noting that we have
\begin{eqnarray}
\sum_{j=1}^{\infty} \sum_{n=2j-1}^{\infty} (\cdot) = \sum_{n=1}^{\infty} \sum_{j=1}^{\mathbb{E} \left( \frac{n+1}{2} \right)} (\cdot) \, ,
\label{double_sum_identity}
\end{eqnarray}
where $\mathbb{E}(x)$ denotes the integer part of the positive real number $x$ (i.e., the natural number $k>0$ such that $k \leqslant x < k + 1$), we find for $m$ odd that
\begin{eqnarray}
Q_{\alpha_1 \cdots \alpha_m \, , \, \beta_1 \cdots \beta_n}^{\mathrm{S}} (\bm{B}) =  \sum_{j=1}^{\mathbb{E} \left( \frac{n+1}{2} \right)}  e_0^{(2j-1)}  \, Q_{\alpha_1 \cdots \alpha_m \, , \, \beta_1 \cdots \beta_n}^{(2j-1)\, \mathrm{S}} (\bm{B})  \, .
\label{coeff_S_Euler}
\end{eqnarray}
This result holds for any odd integer $m \geqslant 1$ and any $n \geqslant 1$. In the case where the index $n$ is even, i.e., $n=2l$ with $l \geqslant 1$, hence making the total number $\mathcal{N} \equiv m+n$ odd, the result~\eref{coeff_S_Euler} is equivalent to equation~(58) of reference~\cite{BG19}. On the other hand, for $n$~odd, i.e., $n=2l-1$ with $l \geqslant 1$, now making the total number $\mathcal{N} \equiv m+n$ even, the result~\eref{coeff_S_Euler} is equivalent to equation~(59) of reference~\cite{BG19}.

Moreover, replacing~\eref{Q_exp_count_par_and_aff}-\eref{A_d_Q} into~\eref{FR_Q+-A-Euler} and using~\eref{Q_k_alt_expr} generates the identity
\begin{eqnarray}
\fl \frac{1}{2} \sum_{m,n=0}^{\infty} \frac{1 + (- 1)^{m}}{m! \, n!} \, Q_{\alpha_1 \cdots \alpha_m \, , \, \beta_1 \cdots \beta_n}^{\mathrm{A}}(\bm{B}) \, \lambda_{\alpha_1} \cdots \lambda_{\alpha_m} A_{\beta_1} \cdots A_{\beta_n} \nonumber \\
\fl = \frac{1}{2} \sum_{m=0}^{\infty} \sum_{j=1}^{\infty} \sum_{n=2j-1}^{\infty} e_0^{(2j-1)} \, \frac{1 - (- 1)^{m+2j-1}}{m! \, n!} \, Q_{\alpha_1 \cdots \alpha_m \, , \, \beta_1 \cdots \beta_n}^{(2j-1) \, \mathrm{A}}(\bm{B}) \, \lambda_{\alpha_1} \cdots \lambda_{\alpha_m} A_{\beta_1} \cdots A_{\beta_n} \, .
\label{Q_plus_A_Q_minus_A}
\end{eqnarray}
Again, identifying the terms with the same powers of $\lambda_{\alpha}$ and $A_{\beta}$ on both sides of~\eref{Q_plus_A_Q_minus_A} and using~\eref{double_sum_identity}, we obtain for $m$ even that
\begin{eqnarray}
Q_{\alpha_1 \cdots \alpha_m \, , \, \beta_1 \cdots \beta_n}^{\mathrm{A}} (\bm{B}) =  \sum_{j=1}^{\mathbb{E} \left( \frac{n+1}{2} \right)}  e_0^{(2j-1)}  \, Q_{\alpha_1 \cdots \alpha_m \, , \, \beta_1 \cdots \beta_n}^{(2j-1)\, \mathrm{A}} (\bm{B})  \, , \label{coeff_A_Euler}
\end{eqnarray}
which is valid for any even integer $m \geqslant 0$ and any $n \geqslant 1$. Now, in the case of an even integer $n=2l$ (with $l \geqslant 1$), hence making the total number $\mathcal{N} \equiv m+n$ even, the result~\eref{coeff_A_Euler} is equivalent to equation~(101) of reference~\cite{BG19}, while for an odd integer $n=2l-1$ (with $l \geqslant 1$), hence yielding an odd $\mathcal{N} \equiv m+n$, the result~\eref{coeff_A_Euler} is equivalent to equation~(102) of reference~\cite{BG19}.

The results~\eref{coeff_S_Euler} and~\eref{coeff_A_Euler} generalize, to systems with a nonzero magnetic field, relations previously obtained in \cite{S92,AG07,AndPhD} in the absence of a magnetic field. Indeed, when $\bm{B}={\bf 0}$ only equation~\eref{coeff_S_Euler} holds, because equation~\eref{coeff_A_Euler} then gives $0=0$. In this case, expressions that can be found in \cite{S92,AG07,AndPhD} are recovered from~\eref{coeff_S_Euler} in view of equations~\eref{e0n-Bernoulli} and~\eref{tanh-Bn}.

\vskip 0.3 cm

\section{Conclusion}\label{conclusion_sec}

In this paper, we investigated the connection between fluctuation relations and Euler's polynomials. We considered a general open system, subjected to an external magnetic field $\bm{B}$, that reaches a nonequilibrium steady state in the long-time limit. The statistical properties of the nonequilibrium currents that take place within the system are then constrained by a fluctuation relation of the form~\eref{fluct_rel}.

The latter is a symmetry property satisfied by the generating function $Q\left( \bm{\lambda} , \bm{A} ; \bm{B} \right)$ of the statistical cumulants, which is a function of the counting parameters $\bm{\lambda}$ and the affinities $\bm{A}$, the magnetic field $\bm{B}$ being treated as a parameter. We reformulated this fluctuation relation in terms of the constant terms $e_0^{(n)}$ of Euler's polynomials $E_n(x)$, with $n \geqslant 1$ an integer that denotes the degree of $E_n(x)$. This could be done by separating the generating function $Q$ into components $Q_{\pm}$ that are even and odd with respect to $\bm{\lambda}$, as well as into symmetric and antisymmetric parts $Q^{\mathrm{S},\mathrm{A}}$ with respect to~$\bm{B}$. Indeed, the fluctuation relation~\eref{fluct_rel} is then mathematically equivalent to the two identities~\eref{FR_Q+-S-Euler} and~\eref{FR_Q+-A-Euler} satisfied by symmetric and antisymmetric parts, respectively. Finally, we showed that these identities yield the sets of relations~\eref{coeff_S_Euler} and~\eref{coeff_A_Euler} for symmetric and antisymmetric quantities, respectively.

A surprising aspect of our recent works \cite{BG18,BG19} has been the use of Euler's polynomials within our mathematical analysis of a fluctuation relation of the form~\eref{fluct_rel}, both in the absence \cite{BG18} and the presence \cite{BG19} of a magnetic field. In particular, we showed in \cite{BG19} that the fluctuation relation constrains about half of the (symmetric and antisymmetric parts of the) cumulants and their responses to the affinities. We then expressed these constrained quantities as linear combinations of the unconstrained ones, the coefficients of which turning to be the constant terms $e_{0}^{(n)}$ of Euler's polynomials. These linear combinations precisely correspond to the relations~\eref{coeff_S_Euler} and~\eref{coeff_A_Euler} inferred in this paper from the alternative expressions~\eref{FR_Q+-S-Euler}-\eref{FR_Q+-A-Euler} of the fluctuation relation~\eref{fluct_rel}. Accordingly, the present work demonstrates that the occurrence of the coefficients of Euler's polynomials happens to be a direct consequence of the symmetry property expressed by fluctuation relations, i.e., microreversibility.


\section*{Acknowledgments}

This research is financially supported by the Universit\'e Libre de Bruxelles (ULB) and the Fonds de la Recherche Scientifique~-~FNRS under the Grant PDR~T.0094.16 for the project ``SYMSTATPHYS".




\section*{References}

\bibliographystyle{unsrt}
\bibliography{Euler4}


\end{document}